\documentclass[10pt,onecolumn,draftcls]{IEEEtran}

\makeatother
\usepackage{tabu}
\usepackage{amsmath}
\usepackage{amsthm}
\usepackage{breqn}
\usepackage{verbatim}
\usepackage{graphicx}
\usepackage{esvect}
\usepackage{cite}
\usepackage{amssymb}
\usepackage{footnote}
\usepackage{color}
\usepackage{url}
\usepackage{subcaption}
\usepackage{xcolor}
\usepackage{transparent}
\usepackage{cancel}
\usepackage{caption}
\usepackage{microtype}
\newtheorem{theorem}{Theorem}[section]

\newtheorem{definition}{Definition}[section]
\newtheorem{lemma}[theorem]{Lemma}
\newtheorem{proposition}{Proposition}[section]
\newtheorem{remark}{Remark}
\newtheorem*{problem statement}{Problem Statement}
\newtheorem{feasibility problem}{Problem}

\newtheorem{assumption}{Assumption}
\definecolor{mygreen}{RGB}{0,108,0}
\definecolor{myred}{RGB}{108,0,0}
\graphicspath{{./Figures/}}
\newcommand{\real}{\mathbb{R}}
\newcommand{\reg}{\mathcal{R}}

\newcommand{\sys}{\mathcal{S}}
\newcommand{\reach}{\mathcal{X}}
\newcommand{\cone}{\mathcal{C}}

\newcommand{\init}{\mathcal{I}}

\newcommand{\treg}{\tilde{\reg}}
\newcommand{\avgratio}{\mathrm{AvgRatio}}
\newcommand{\avgdiff}{\mathrm{AvgDiff}}

\begin{document}
	\title{Abstracting the Traffic of Nonlinear Event-Triggered Control Systems} 
	\author{Giannis Delimpaltadakis and Manuel Mazo Jr.\thanks{The authors are with the Delft University of Technology, The Netherlands. Emails:\{i.delimpaltadakis, m.mazo\}@tudelft.nl. This work is supported by the ERC Starting Grant SENTIENT (755953).}}
	\maketitle
	\graphicspath{{./Figures/}}
	\begin{abstract}
		Scheduling communication traffic in networks of event-triggered control (ETC) systems is challenging, as their sampling times are unknown, hindering application of ETC in networks. In previous work, finite-state abstractions were created, capturing the sampling behaviour of LTI ETC systems with quadratic triggering functions. Offering an infinite-horizon look to all sampling patterns of an ETC system, such abstractions can be used for scheduling of ETC traffic. Here we significantly extend this framework, by abstracting perturbed uncertain nonlinear ETC systems with general triggering functions. To construct an ETC system's abstraction: a) the state space is partitioned into regions, b) for each region an interval is determined, containing all intersampling times of points in the region, and c) the abstraction's transitions are determined through reachability analysis. To determine intervals and transitions, we devise algorithms based on reachability analysis. For partitioning, we propose an approach based on isochronous manifolds, resulting into tighter intervals and providing control over them, thus containing the abstraction's non-determinism. Simulations showcase our developments.
	\end{abstract}
	
	\section{Introduction}
	Event-Triggered Control (ETC, \cite{astrom2002comparison, tabuada2007etc, small_gain_robust_etc, girard2015dynamicetc, 2012introtoetc_stc}) determines sampling instants such that communication between the sensors and the controller is efficient, while certain performance specifications are met (e.g. stability). The sensors measure continuously the system's state, and transmit measurements only when they detect satisfaction of a certain \textit{triggering condition}.

	Although the vast research on ETC shows promising results on reducing bandwidth/energy usage, there are open problems, hindering its application in shared networks. One such problem is scheduling of communication traffic in networks of multiple ETC loops; i.e., determining at each time which of the loops will occupy the communication channel, such that they all communicate timely without packet collisions, while the desired performance is met. In contrast to periodic sampling, where sampling instants are known by construction, ETC sampling times are generally unknown, due to the sampling's event-based nature. This renders scheduling of ETC traffic a challenging problem. One way of approaching it is the co-design techniques of \cite{buttazzo1998elastic, caccamo2000elastic, bhattacharya2004anytime, fontanelli2008scheduling, areqi2015scheduling, lu2002feedback, cervin2005control}. According to such strategies, given a network of control loops, the controllers, the sampling instants and the scheduler are designed in a coupled manner. 
	However, these approaches lack versatility, as the whole design process is applied from scratch, when a new loop joins the network.
	
	A different approach is based on \textit{abstractions} \cite{arman_formal_etc}-\hspace{-.1mm}\cite{gleizer2020scalable}. According to it, ETC systems are abstracted by finite-state \textit{quotient systems} (abstractions), capturing the ETC systems' sampling behaviour. The abstraction's set of output sequences contains all possible sequences of intersampling times that the given ETC system may exhibit, thus providing an infinite-horizon look into its sampling patterns. Employing this property, \cite{arman_formal_etc} and \cite{gleizer2020scalable} showed that such abstractions can be employed for scheduling of ETC traffic. This approach is more versatile compared to the co-design techniques, as the abstraction of each system in the network is computed only once offline, and does not change with the presence of a new system. 
	
	To construct the abstraction, the system's state-space is partitioned into finitely many regions $\treg_{i,j}$, representing the abstraction's states. For each region $\treg_{i,j}$, an interval $[\underline{\tau}_{\treg_{i,j}}, \overline{\tau}_{\treg_{i,j}}]$ is determined, containing all intersampling times corresponding to states in the region. These intervals serve as the abstraction's output. Finally, the abstraction's transitions are determined via reachability analysis (e.g. see \cite{dreach, flowstar}). The abstraction's non-determinism, encoding how coarsely it captures the actual system's behaviours, depends on the intervals' tightness and the transition set's size. Previous works \cite{arman_formal_etc}-\hspace{-.1mm}\cite{gleizer2020scalable} abstracted LTI systems with quadratic triggering functions.
	
	Here, we significantly extend the above framework by abstracting the traffic of nonlinear ETC systems with disturbances, uncertainties and general triggering functions. To determine the timing intervals and the transitions, we propose an algorithm based on reachability analysis. Regarding state-space partitioning, we propose an approach that is based on approximations of \textit{isochronous manifolds} (IMs, sets in the state-space with uniform intersampling time), previously derived in \cite{delimpaltadakis_tac} and \cite{delimpaltadakis2020region}. By partially inheriting the merits of partitioning with actual IMs, this approach aims at providing control over the timing intervals and improving their tightness, thus containing one source of the abstraction's non-determinism. Simulation comparisons between the proposed partition and a naive partition support our arguments, as the proposed partition achieves tighter intervals (for metrics capturing tightness refer to Section \ref{section_numerical}).
	Finally, we note that a preliminary version of the present article, focusing only on homogeneous systems and triggering functions, has been presented in \cite{delimpaltadakis2020traffic}. 
	
	To summarize our contributions, in this work we:
		\begin{itemize}
			\item construct traffic abstractions of perturbed uncertain nonlinear ETC systems with general triggering functions, thus significantly extending abstraction-based scheduling of ETC traffic, which was only applicable to LTI systems with quadratic triggering functions so far (\hspace{-.1mm}\cite{arman_formal_etc,gleizer2020scalable}),
			\item formulate and solve reachability analysis problems, providing intervals containing the intersampling times of all points in a given state-space region, 
			\item propose a state-space partition, that provides control over the timing intervals and improves their tightness, thus containing a source of the abstraction's non-determinism.
	\end{itemize}
	
	\section{Notation and Preliminaries}
	\subsection{Notation}
	The Euclidean norm of a point $x\in\real^n$ is denoted by $|x|$. For vectors, we also use the notation $(x_1,x_2)=\begin{bmatrix}x_1^\top &x_2^\top\end{bmatrix}^\top$. For a set $X\subseteq \real^n$, $2^X$ denotes its power-set. Given two subsets $X_a,X_b\subseteq X$, $d_H(X_a,X_b)$ denotes their Hausdorff distance. Given an equivalence relation $Q\subseteq X\times X$, the set of all equivalence classes is denoted by $X/Q$. 
	
	Consider the system of ordinary differential equations:
	\begin{equation}\label{ode}
	\dot{\zeta}(t) =f(\zeta(t)),
	\end{equation}
	where $\zeta(t)\in\real^n$ and $f:\real^n\to\real^n$. A solution to \eqref{ode} with initial condition $\zeta_0$ and initial time $t_0$ is denoted by $\zeta(t;t_0,\zeta_0)$. When $t_0$ (and $\zeta_0$) is clear from the context, we omit it by writing $\zeta(t;\zeta_0)$ (respectively $\zeta(t)$). Given a set of initial states $\init\subseteq\real^n$, the \textit{reachable set} of \eqref{ode} at time $T$ is $\reach^f_T(\init) := \{\zeta(T;\zeta_0):\zeta_0\in\init\}$. 
	Likewise, the \textit{reachable flowpipe} of \eqref{ode} in the time interval $[\tau_1,\tau_2]$, with initial set $\init$, is $\reach^f_{[\tau_1,\tau_2]}(\init) := \bigcup\limits_{T\in[\tau_1,\tau_2]}\reach^f_T(\init).$

	\subsection{Systems and Simulation Relations}
	Here we recall notions of systems and simulation relations from \cite{tabuada_book}, which are employed later.
	\begin{definition}[System {\cite[Definition 1.1]{tabuada_book}}]
		A system $\mathcal{S}$ is a tuple $(X, X_0, $ $\longrightarrow, Y, H)$, where $X$ is the set of states, $X_0$ the set of initial states, $\longrightarrow \subseteq X\times X$ a transition relation, $Y$ the set of outputs and $H:X\to Y$ the output map.
	\end{definition}
	We have omitted the action set $U$ from the above definition, since we only focus on autonomous systems. If $X$ is a finite (or infinite) set, then $\mathcal{S}$ is called finite-state (respectively infinite-state). A system $\mathcal{S}$ is called a \textit{metric system} if $Y$ is equipped with a metric $d:Y\times Y\to\real_0^+\cup\{+\infty\}$.
	\begin{definition}[$\epsilon$-Approximate Simulation Relation {\cite[Definition 9.2]{tabuada_book}}]\label{def_epsilon_approximate}
		Consider two metric systems $\sys_a,\sys_b$ with $Y_a=Y_b$ and a constant $\epsilon\geq0$. A relation $Q\subseteq X_a\times X_b$ is an $\epsilon$-approximate simulation relation from $\sys_a$ to $\sys_b$ if it satisfies:
		\begin{itemize}
			\item $\forall x_{0_a}\in X_{0_a}: \quad \exists x_{0_b}\in X_{0_b}$ such that $(x_{0_a},x_{0_b})\in Q$,
			\item $\forall (x_a,x_b)\in Q: \quad d(H_a(x_a), H_b(x_b))\leq\epsilon$,
			\item $\forall x_a,x_a' \in X_a$ with $(x_a, x_a')\in \underset{a}{\longrightarrow}:$ if $(x_a,x_b)\in Q$ then $\exists (x_b, x_b')\in \underset{b}{\longrightarrow}$ such that $(x_a',x_b')\in Q$.
		\end{itemize}
	\end{definition}
	If there exists an $\epsilon$-approximate simulation relation from $\sys_a$ to $\sys_b$, we say that $\sys_b$ $\epsilon$-approximately simulates $\sys_a$ and write $\sys_a \overset{\epsilon}{\preceq}\sys_b$. Moreover, let us introduce an alternative definition of \textit{power quotient systems}. For the original definition, see \cite{tabuada_book}.
	\begin{definition}[Power Quotient System {\cite[Definition 6]{arman_formal_etc}}]
		Consider a system $\sys = (X, X_0, \longrightarrow, Y, H)$ and an equivalence relation $Q\subseteq X\times X$. The power quotient system of $\sys$ is the tuple $\sys_{/ Q}= (X_{/ Q}, X_{0_{/ Q}}, \underset{{/ Q}}{\longrightarrow}, Y_{/ Q}, H_{/ Q})$, where:
		\begin{itemize}
			\item $X_{/ Q} = X/ Q$ and $X_{0_{/ Q}} = \{x_{/ Q} \in X_{/ Q}:x_{/ Q}\cap X_0 \neq \emptyset \}$,
			\item $(x_{/ Q}, x'_{/ Q})\in\underset{{/ Q}}{\longrightarrow}$ if $\text{ }\exists (x,x')\in \longrightarrow$ such that $x \in x_{/ Q}$ and $x' \in x'_{/ Q}$,
			\item $Y_{/ Q}\subseteq2^Y$ and $H_{/ Q}(x_{/ Q})=\bigcup\limits_{x\in x_{/ Q}}H(x)$.
		\end{itemize}
	\end{definition}
	\begin{lemma}[\hspace{1sp}{\cite[Lemma 1]{arman_formal_etc}}] \label{precision lemma}
		Consider a metric system $\sys$, a relation $Q\subseteq X\times X$ and the power quotient system $\sys_{/ Q}$. For any $\epsilon$ such that $\epsilon \geq \sup\limits_{x\in x_{/Q},\text{ }x_{/Q}\in X/Q}d_H(H(x),H_{/ Q}(x_{/ Q})),$ 
		$\sys_{/ Q}$ $\epsilon$-approximately simulates $\sys$, i.e. $\sys \overset{\epsilon}{\preceq}\sys_{/ Q}$.
	\end{lemma}
	
	\subsection{Event-Triggered Control Systems}
	Consider the following control system with state feedback:
	\begin{equation} \label{ct_sys}
	\dot{\zeta}(t) = f\Big(\zeta(t),\upsilon(\zeta(t))\Big),
	\end{equation}
	where $\zeta:\real_0^+\to\real^n$, $f:\real^n\to\real^n$ and $\upsilon:\real^n\to\real^m$. In a sample-and-hold digital implementation of \eqref{ct_sys}, the input is held constant between consecutive \textit{sampling time instants} $t_i$ and is only updated at sampling times, i.e.:
	\begin{equation}\label{snh}
	\dot{\zeta}(t) = f\Big(\zeta(t),\upsilon(\zeta(t_i))\Big), \quad t\in[t_i,t_{i+1}).
	\end{equation}
	The so-called \textit{sampling-induced error} is the deviation of the current state of \eqref{snh} from the last measurement:
	\begin{equation*}
	\varepsilon_\zeta(t) = \zeta(t_i)-\zeta(t), \quad t\in[t_i,t_{i+1}).
	\end{equation*}
	Observe that $\varepsilon_\zeta(t)$ resets to zero, at each sampling time $t_i$. By employing $\varepsilon_\zeta(t)$, we can write \eqref{snh} as:
	\begin{equation}\label{snh2}
	\dot{\zeta}(t) = f\Big(\zeta(t),\upsilon(\zeta(t)+\varepsilon_\zeta(t))\Big), \quad t\in[t_i,t_{i+1}).
	\end{equation}
	In ETC, sampling times are defined as follows:
	\begin{equation}\label{trig_cond}
	t_{i+1} = t_i + \inf\{t>0:\text{ }\phi(\zeta(t;x_i),\varepsilon_{\zeta}(t))> 0\}
	\end{equation}
	and $t_0=0$, where  $x_i=\zeta(t_i)$ is the last state measurement, $\phi(\cdot,\cdot)$ is the \textit{triggering function}, \eqref{trig_cond} is the \textit{triggering condition}, and $t_{i+1}-t_i$ is called \textit{intersampling time}. Each point $x\in\real^n$ is associated to a unique intersampling time $\tau(x)$:
	\begin{equation}\label{intersampling_time}
	\tau(x):=\inf\{t>0:\text{ }\phi(\zeta(t;x),\varepsilon_{\zeta}(t))> 0\}.
	\end{equation}
	Between two sampling times $t_i$ and $t_{i+1}$, the triggering function starts from a negative value $\phi(\zeta(t_i),0)<0$ ($\varepsilon_\zeta$ is zero at sampling times) and remains negative until it becomes zero at $t_{i+1}^-$. At $t_{i+1}^+$, the state is sampled again, the sampling-induced error resets to zero, the triggering function resets to a negative value and the control action is updated.
	
	By observing that $\dot{\varepsilon}_{\zeta}(t)=-\dot{\zeta}(t)$, we write the dynamics of the ETC system in the following extended form:
	\begin{equation} \label{etc_system}
	\begin{aligned}
	&\dot{\xi}(t)= \begin{bmatrix} f\Big(\zeta(t),\upsilon(\zeta(t)+\varepsilon_{\zeta}(t))\Big)\\
	-f\Big(\zeta(t),\upsilon(\zeta(t)+\varepsilon_{\zeta}(t))\Big) \end{bmatrix}=:f_e(\xi(t)), \text{ } t \in [t_i, t_{i+1})\\
	&\xi(t_{i+1}^+)=\begin{bmatrix}
	\zeta^\top(t^+_{i+1}) &0^\top
	\end{bmatrix}^\top,
	\end{aligned}
	\end{equation}
	where $\xi = (\zeta,\varepsilon_{\zeta})\in\real^{2n}$. At each sampling time $t_i$, the state of \eqref{etc_system} becomes $\xi_i=(x_i,0)$. Thus, since we focus on intervals between consecutive sampling times $[t_i,t_{i+1})$, instead of writing $\phi\Big(\xi(t;(x_i,0))\Big)$ (or $\tau\Big((x_i,0)\Big)$), we abusively write $\phi(\xi(t;x_i))$ (or $\tau(x_i)$) for convenience.
	
	\section{Problem Statement}
	In this work, we construct \emph{traffic abstractions} of nonlinear ETC systems; we construct finite-state systems, whose set of output sequences contains all possible intersampling time sequences of the given ETC system. For clarity, we mainly consider the case without disturbances or uncertainties, but we also point out through remarks (Remarks \ref{remark_perturbed_reach} and \ref{remark_perturbed_partition}) how the proposed approach directly applies to disturbances/uncertainties.
	
	We adopt a problem formulation similar to \cite{arman_formal_etc}. Consider the ETC system \eqref{snh2}-\eqref{trig_cond}. Let us introduce the system:
	\begin{equation}\label{unperturbed_sys}
	\sys = (X,X_0,\longrightarrow,Y,H),
	\end{equation}
	where $X=X_0\subseteq\real^n$, $Y\subseteq\real^+$, $H(x)=\tau(x)$ and the transition relation $\longrightarrow\subseteq X\times X$ is such that $(x,x')\in\longrightarrow$ $\iff$ $\zeta(\tau(x);x)=x'$. Observe that the set of output sequences of system \eqref{unperturbed_sys} contains all possible intersampling time sequences of the ETC system \eqref{snh2}-\eqref{trig_cond}; that is, system \eqref{unperturbed_sys} captures exactly the traffic of the ETC system. However, it is infinite-state and cannot serve as a computationally handleable abstraction.
	
	We, also, introduce the following set of assumptions:
	\begin{assumption} \label{assum_1}
		\hfill
		\begin{enumerate}
			\item The origin is the only equilibrium of \eqref{ct_sys}.\label{assum_1_origin}
			\item The vector field $f_e(\cdot)$ from \eqref{etc_system} is locally bounded.\label{assum_1_vector_field_bounded}
			\item $\phi(0,0)\leq 0$ and $\phi(x,0)<0$ for all $x\neq0$. Moreover, for any compact set $K\subset\real^n$ there exists $\epsilon_K>0$ such that for all $x_0\in K$, $\phi(\xi(t;x_0))\leq0$ for all $t\in[0,\epsilon_K)$.\label{assum_1_tri_fun_2}
			\item The set $X$ is compact and connected. \label{assum_1_X}
		\end{enumerate}
	\end{assumption}
	Item \ref{assum_1_origin} serves for clarity of presentation. Item \ref{assum_1_tri_fun_2} imposes that $\phi(\cdot,0)$ is negative-definite and that the given ETC system cannot exhibit infinitely fast sampling; this is satisfied by most functions in the ETC literature (e.g. Tabuada's \cite{tabuada2007etc}, dynamic triggering \cite{girard2015dynamicetc}, mixed triggering \cite{small_gain_robust_etc}, Lebesgue sampling \cite{astrom2002comparison}). 
	Item \ref{assum_1_X} suggests that we are interested in trajectories of the system that stay in the compact connected set $X$.
	
	Since \eqref{unperturbed_sys} captures exactly the sampling behaviour of the ETC system \eqref{snh2}-\eqref{trig_cond}, abstracting the ETC system is equivalent to abstracting \eqref{unperturbed_sys}. This gives rise to the following:
	\begin{problem statement}\label{prob_state_1}
		Consider the system $\sys$ \eqref{unperturbed_sys}. Let Assumption \ref{assum_1} hold. Construct a power-quotient system $\sys_{/ Q}= (X_{/ Q}, X_{0_{/ Q}}, \underset{{/ Q}}{\longrightarrow}, Y_{/ Q}, H_{/ Q})$ with:
		\begin{itemize}
			\item $X_{/ Q} = X/Q := \{\treg_{1,1},\dots,\treg_{i,j},\dots,\treg_{q,m}\}$ and $X_{0_{/ Q}} = X_{/ Q}$, where $\treg_{i,j}\subseteq X$ and $\bigcup\treg_{i,j}=X$.
			\item $(x_{/Q},x'_{/Q})\in\underset{{/ Q}}{\longrightarrow}$ if $\exists x \in x_{/Q}$ and $\exists x' \in x'_{/Q}$ such that $\zeta(H(x);x)=x'$,
			\item $Y_{/Q}\subseteq 2^Y = 2^{\real^+}$ and $H_{/ Q}(\treg_{i,j}) := [\underline{\tau}_{\treg_{i,j}},\overline{\tau}_{\treg_{i,j}}]$, with:
			\begin{equation}\label{intervals}
			\underline{\tau}_{\treg_{i,j}} \leq \inf\limits_{x\in \treg_{i,j}}H(x), \quad \overline{\tau}_{\treg_{i,j}} \geq \sup\limits_{x\in \treg_{i,j}}H(x).
			\end{equation}
		\end{itemize}
	\end{problem statement}
	The states $\treg_{i,j}$ of the abstraction are regions in the ETC system's state-space, i.e. $\treg_{i,j}\subseteq X\subset\real^n$ (the $i,j$-subscript becomes clear later). A transition from $\treg_{i,j}$ to $\treg_{k,l}$ is defined if there exists a trajectory starting from $x\in\treg_{i,j}$, which ends up in $\treg_{k,l}$ after an elapsed intersampling time $\tau(x)$. Hence, a transition is taken every time the triggering condition \eqref{trig_cond} is satisfied. Finally, \eqref{intervals} indicates that the abstraction's output of a state $\treg_{i,j}$ is an interval containing all intersampling times corresponding to states $x\in\treg_{i,j}$. Thus, given a run of the ETC system, there is a corresponding run of the abstraction, whose output sequence is a sequence of intervals containing the intersampling time that the ETC system exhibited at that particular step of the run. In fact, by Lemma \ref{precision lemma}, we conclude that $\sys \overset{\epsilon}{\preceq}\sys_{/ Q}$ for all $\epsilon\geq\max\limits_i\{\overline{\tau}_{\treg_{i,j}}-\underline{\tau}_{\treg_{i,j}}\}$.
	
	As discussed in \cite{arman_formal_etc}, the abstraction $\sys_{/Q}$ is semantically equivalent to a \emph{timed-automaton}. The automaton's guards are determined by the intervals $[\underline{\tau}_{\treg_{i,j}},\overline{\tau}_{\treg_{i,j}}]$, and its transitions are the ones of $\sys_{/Q}$. The tighter the intervals and the smaller the transition set, the less non-deterministic becomes the automaton; hence it simulates more accurately the original system, and the scheduling algorithms provide less conservative results. 
	
	Finally, to address the problem, we need to partition $X$ into regions $\treg_{i,j}$ (which automatically generates the relation $Q$), derive the timing intervals, and determine the transitions. In Section \ref{section_reach}, we propose reachability-analysis-based algorithms to determine  the timing intervals and transitions, given any partition. In Section \ref{section_partition}, we propose a partition, providing better control over the intervals and their tightness, thus containing one of the sources of the abstraction's non-determinism.
	
	\section{Timing Intervals and Transitions} \label{section_reach}
	In this section, we assume that the partition is given and show how reachability analysis can be employed to determine timing intervals and transitions.
	
	\subsection{Reachability Analysis for Timing Intervals}\label{section_reach_intervals}
		The following proposition, employing reachable sets and flowpipes, provides conditions that determine lower and upper bounds on intersampling times of points in a given region $\treg_{i,j}$:
		\begin{proposition}\label{reach_intervals_prop}
			Consider the ETC system \eqref{snh2}-\eqref{trig_cond} and its extended form \eqref{etc_system}. Let Assumption \ref{assum_1} hold. Let $\treg_{i,j}\subseteq X$. Define the following sets:
			\begin{align*}
			\init_{i,j}&:=\{(x,0)\in\real^{2n}:x\in\treg_{i,j}\}\\
			\mathcal{U}_{> 0}&:=\{(x,e)\in\real^{2n}:\phi\Big((x,e)\Big)> 0\}\\
			\mathcal{U}_{\leq 0}&:=\{(x,e)\in\real^{2n}:\phi\Big((x,e)\Big)\leq 0\}
			\end{align*}
			If:
			\begin{equation}\label{lower bound cond}
			\reach^{f_e}_{[0,\tau_{\min}]}(\init_{i,j})\cap\mathcal{U}_{>0}=\emptyset,
			\end{equation}then for all $x\in\treg_{i,j}$: $\tau(x)\geq\tau_{\min}$,
			where $\tau(\cdot)$ is as in \eqref{intersampling_time}. Similarly, if:
			\begin{equation}\label{upper bound cond}
			\reach^{f_e}_{\tau_{\max}}(\init_{i,j})\cap\mathcal{U}_{\leq 0}=\emptyset,
			\end{equation}then for all $x\in\treg_{i,j}$: $\tau(x)\leq\tau_{\max}$.
		\end{proposition}
		\begin{proof}
			Equation \eqref{lower bound cond} implies that $\forall x\in\treg_{i,j}$ we have that: $\phi(\xi(t;x))\leq0$, for all $t\in [0,\tau_{\min}]$. Thus, $\tau(x)\geq\tau_{\min}$, i.e. $\tau_{\min}$ is a lower bound on intersampling times of region $\treg_{i,j}$.
			
			Similarly, if $\reach^{f_e}_{\tau_{\max}}(\init_{i,j})\cap\mathcal{U}_{\leq0}=\emptyset$, then for all $x\in\treg_{i,j}$ we have that $\phi(\xi(\tau_{\max};x))>0$. Thus, $\tau(x)\leq\tau_{\max}$.
		\end{proof}
		
		To obtain the timing intervals $[\underline{\tau}_{\treg_{i,j}},\overline{\tau}_{\treg_{i,j}}]$ for regions $\treg_{i,j}$, we employ a line search on the variables $\tau_{\min},\tau_{\max}$ and iterate until we find that \eqref{lower bound cond} and \eqref{upper bound cond} are satisfied. To check \eqref{lower bound cond} and \eqref{upper bound cond}, we employ reachability-analysis computational tools (e.g. \cite{flowstar,dreach}). Such tools, given a system \eqref{ode}, a set of initial conditions $\init\subset\real^n$ and a set $\mathcal{U}\subseteq\real^n$, overapproximate reachable flowpipes $\reach^f_{[\tau_1,\tau_2]}(\init)$ and the set $\mathcal{U}$ by overapproximations $\hat{\reach}^f_{[\tau_1,\tau_2]}(\init)\supseteq\reach^f_{[\tau_1,\tau_2]}(\init)$ and $\hat{\mathcal{U}}\supseteq\mathcal{U}$, and check if $\hat{\reach}^f_{[\tau_1,\tau_2]}(\init)\cap\hat{\mathcal{U}}=\emptyset$. Moreover, by the implication:
		\begin{align}
		&\hat{\reach}^f_{[\tau_1,\tau_2]}(\init)\cap\hat{\mathcal{U}}=\emptyset \implies \reach^f_{[\tau_1,\tau_2]}(\init)\cap\mathcal{U}=\emptyset,\label{overapprox_reach}
		\end{align}
		they can determine if $\reach^f_{[\tau_1,\tau_2]}(\init)\cap\mathcal{U}=\emptyset$. Hence, by employing a line search on $\tau_{\min}$ and $\tau_{\max}$, via a reachability analysis tool we check iteratively if $\hat{\reach}^{f_e}_{[0,\tau_{\min}]}(\init_{i,j})\cap\hat{\mathcal{U}}_{>0}=\emptyset$ and $\hat{\reach}^{f_e}_{\tau_{\max}}(\init_{i,j})\cap\hat{\mathcal{U}}_{\leq 0}=\emptyset$, until these conditions are satisfied. Satisfaction of these conditions implies satisfaction of \eqref{lower bound cond} and \eqref{upper bound cond} (due to \eqref{overapprox_reach}), which implies that $\tau(x)\geq\tau_{\min}=\underline{\tau}_{\treg_{i,j}}$ $\tau(x)\leq\tau_{\max}=\overline{\tau}_{\treg_{i,j}}$ for all $x\in\treg_{i,j}$, by Proposition \ref{reach_intervals_prop}.
		
		\begin{remark}\label{heartbeat_remark}
				Certain regions might not admit upper bounds on their intersampling times (e.g. in equilibria $x$, $\tau(x)=+\infty$). In practice, to cope with this, an arbitrary maximum intersampling time $\tau_h$ is introduced (called ``heartbeat"), such that sampling instants are determined by $t_{i+1}=t_i + \min(\tau(x_i),\tau_h)$, where $x_i$ is the last measured state and $\tau(\cdot)$ is defined in \eqref{intersampling_time}. Thus, for such regions $\treg_{i,j}$, we can arbitrarily dictate an upper bound $\overline{\tau}_{\treg_{i,j}}=\tau_h$ to be equal to the heartbeat, and force the sensors to sample according to $t_{i+1}=t_i + \min(\tau(x_i),\tau_h)$.
		\end{remark}
	
	\subsection{Reachability Analysis for Transitions}
	Now, let us show how reachability analysis can be used to derive the abstraction's transitions. Recall the transitions' definition, from Problem Statement \ref{prob_state_1}:
	\begin{equation*}
	(\treg_{i,j},\treg_{k,l})\in\underset{{/ Q}}{\longrightarrow}\text{, if }:
	\end{equation*}
	\begin{equation*}
	\exists x \in \treg_{i,j} \text{ and } \exists x' \in \treg_{k,l} \text{ such that } \zeta(H(x);x)=x'.
	\end{equation*}
	This definition can be relaxed as follows:
	\begin{equation}\label{transition_def}
	(\treg_{i,j},\treg_{k,l})\in\underset{{/ Q}}{\longrightarrow}\text{, if }: \text{ }\reach^f_{[\underline{\tau}_{\treg_{i,j}},\overline{\tau}_{\treg_{i,j}}]}(\treg_{i,j})\cap\treg_{k,l}\neq\emptyset. 
	\end{equation}
	Thus, inspired by \eqref{overapprox_reach}, via a reachability analysis tool we check if $\hat{\reach}^f_{[\underline{\tau}_{\treg_{i,j}},\overline{\tau}_{\treg_{i,j}}]}(\treg_{i,j})\cap\treg_{k,l}\neq\emptyset$, which approximates condition \eqref{transition_def}, and if satisfied we define a transition $(\treg_{i,j},\treg_{k,l})\in\underset{{/ Q}}{\longrightarrow}$.
	
	In this way, the constructed abstraction contains all possible transitions $(\treg_{i,j},\treg_{k,l})$  defined as in \eqref{transition_def}. Notice that, since \eqref{transition_def} is a relaxation of the original transitions' definition, and $\hat{\reach}^f_{[\underline{\tau}_{\treg_{i,j}},\overline{\tau}_{\treg_{i,j}}]}(\treg_{i,j})\cap\treg_{k,l}\neq\emptyset$ does not necessarily imply that $\reach^f_{[\underline{\tau}_{\treg_{i,j}},\overline{\tau}_{\treg_{i,j}}]}(\treg_{i,j})\cap\treg_{k,l}\neq\emptyset$, the abstraction may contain additional transitions $(\treg_{i,j},\treg_{k,l})$ for which $\not\exists x\in\treg_{i,j}$ and $\not\exists x'\in\treg_{k,l}$ such that $\zeta(H(x);x)=x'$. Nonetheless, the existence of spurious transitions does not affect the fact that $\sys_{/Q}$ $\epsilon$-approximately simulates $\sys$ (see \cite{tabuada_book}).
		\begin{remark}\label{remark_tradeoff}
			Since reachability analysis uses overapproximations, the computed intervals and transitions are not exact. Nonetheless, higher accuracy settings for reachability analysis imply more accurate intervals and transitions, establishing a trade-off between accuracy and offline computations.
	\end{remark}
	\begin{remark}
		Overapproximations $\hat{\reach}^{f}_{[\underline{\tau}_{\treg_{i,j}},\overline{\tau}_{\treg_{i,j}}]}(\treg_{i,j})$ of the flowpipes of the ETC system \eqref{snh2} can be readily obtained by the -already computed from the previous step- flowpipes $\hat{\reach}^{f_e}_{[\underline{\tau}_{\treg_{i,j}},\overline{\tau}_{\treg_{i,j}}]}(\init_{i,j})$ of the extended system \eqref{etc_system}, by projecting to the $\zeta$-variables: $\hat{\reach}^{f}_{[\underline{\tau}_{\treg_{i,j}},\overline{\tau}_{\treg_{i,j}}]}(\treg_{i,j}) = \boldsymbol{\pi}_\zeta\hat{\reach}^{f_e}_{[\underline{\tau}_{\treg_{i,j}},\overline{\tau}_{\treg_{i,j}}]}(\init_{i,j})$. Thus, the only computation needed to determine transitions is calculating the intersections $\boldsymbol{\pi}_\zeta\hat{\reach}^{f_e}_{[\underline{\tau}_{\treg_{i,j}},\overline{\tau}_{\treg_{i,j}}]}(\init_{i,j})\cap \treg_{k,l}$. This is in contrast to \cite{arman_formal_etc}, where computing timing intervals and determining transitions are two distinct computational steps. 
	\end{remark}
	\begin{remark}\label{remark_perturbed_reach}
		The above method directly extends to systems with bounded disturbances/uncertainties, since
		many reachability analysis tools, such as Flow* \cite{flowstar}, can handle bounded unknown signals.
	\end{remark}
	
	\section{Partitioning the State Space}\label{section_partition}
	Here, we propose a way of partitioning the state space into regions $\treg_{i,j}$, based on approximations of isochronous manifolds (IMs), derived in \cite{delimpaltadakis_tac} and \cite{delimpaltadakis2020region}, providing control over the timing intervals and improving their tightness, compared to naively partitioning $X$ into polytopes. First, we present the ideal (albeit non-achievable) partitioning in these terms, which employs IMs. Afterwards, we show how to approximate it via inner approximations of IMs: we start with homogeneous ETC systems, and then we generalize employing a homogenization procedure. Finally, we provide a thorough discussion on the advantages of the proposed approach. For this section, we add the following mild assumptions:
	\begin{assumption}\label{assum_2} The vector field $f_e(\cdot)$ of \eqref{etc_system} is continuous. The function $\phi(\cdot)$ is continuously differentiable.
	\end{assumption}
	
	\subsection{Isochronous Manifolds and Ideal Partitioning}\label{section_IM_partition}
	Here, we demonstrate how IMs, if obtained exactly, enable a partition (hereby termed IM-partition) which is ideal w.r.t. the timing intervals: it a) provides complete control over the intervals, and b) is optimal in terms of correspondence between timing intervals and state-space regions. We focus on homogeneous systems and triggering functions, for clarity.
	
\begin{figure}[!h]
		\centering
		\includegraphics[width=1.4in]{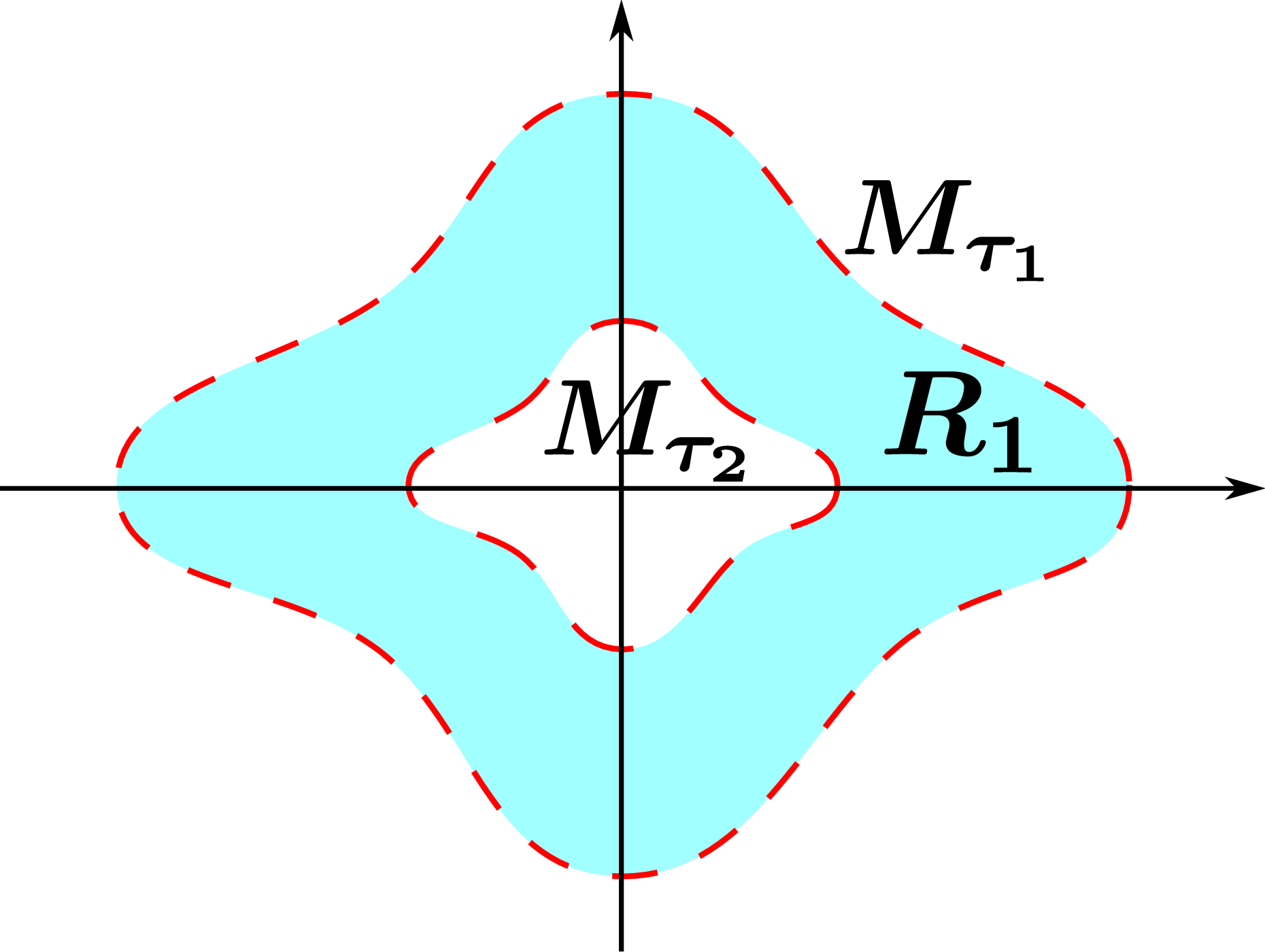}
		\caption{IMs (dashed lines) of a homogeneous ETC system for times $\tau_1<\tau_2$. The region $R_1$ (filled region) satisfies \eqref{reg_between_manfolds_interval}.}
		\label{fig_two_manifolds}
	\end{figure}

	\begin{definition}[Homogeneous function {\cite[Definition IV.1, simplified]{delimpaltadakis_tac}}]
		Consider a function $f:\real^n\to\real^m$. We say that $f$ is homogeneous of degree $\alpha\in\real$, if for all $x\in\real^n$ and any $\lambda>0$: $f(\lambda x) = \lambda^{\alpha+1}f(x)$.
	\end{definition}
	A dynamical system \eqref{ode} is called homogeneous of degree $\alpha\in\real$ if $f$ is homogeneous of the same degree.
	\begin{definition}[Isochronous Manifold {\cite[Definition IV.3]{delimpaltadakis_tac}}]
		Consider an ETC system \eqref{snh2}-\eqref{trig_cond}. The set $M_{\tau_{\star}}=\{x\in\mathbb{R}^n : \tau(x)=\tau_{\star}\}$, where $\tau(x)$ is as in \eqref{intersampling_time}, is called isochronous manifold of time $\tau_{\star}$.
		\label{manifold definition}
	\end{definition}
	It becomes clear how IMs constitute a notion relating regions in a system's state-space and intersampling times: they are sets of points in the state-space, with the same intersampling time. As discussed in \cite{delimpaltadakis_tac} and \cite{delimpaltadakis2020region}, for homogeneous ETC systems and triggering functions, IMs satisfy certain useful properties (e.g. listed in \cite[Proposition IV.3]{delimpaltadakis2020region}). Due to these properties, the sets $R_i$ consisting of the points lying between two manifolds of times $\tau_i,\tau_{j}$ with $\tau_i\leq\tau_{j}$ (see Fig. \ref{fig_two_manifolds}) satisfy:
	\begin{equation}\label{reg_between_manfolds_interval}
	R_i = \{x\in\real^n:\tau(x)\in[\tau_i,\tau_j]\},
	\end{equation}
	i.e. $R_i$ is the set of all points with intersampling times in $[\tau_i,\tau_j]$. Thus, if IMs were obtained exactly, one could: choose a set of times $\{\tau_1,\tau_2,...,\tau_q\}$, generate the IMs $M_{\tau_i}$, and use the regions $R_i$ between successive IMs to partition the state-space. 
	
	The advantages of IM-partitioning are the following. First, complete control over the timing intervals is obtained, as the regions $R_i$ are generated such that the corresponding timing intervals are equal to the chosen ones $[\tau_i,\tau_{i+1}]$ (due to \eqref{reg_between_manfolds_interval}). Moreover, the IM-partition is optimal w.r.t. correspondence between regions and intervals: due to \eqref{reg_between_manfolds_interval}, there is no set with bigger volume (Lebesgue measure) than $R_i$ that corresponds to the same timing interval. This implies that IM-partitioning achieves the tightest intervals possible than any other partition, given a certain volume (or number) of regions. 
	
	Unfortunately, IMs cannot be obtained exactly for most systems. However, the above advantages of IM-partitioning motivate us to employ inner-approximations of IMs (\hspace{-.1mm}\cite{delimpaltadakis_tac,delimpaltadakis2020region}), in order to approximate this ideal way of partitioning.
	
	\subsection{State Space Partitioning for Homogeneous Systems via Inner Approximations of IMs}
	For clarity of presentation, we first present how inner-approximations of IMs can be employed to partition the state space of homogeneous systems and triggering functions (recalled from the preliminary version \cite{delimpaltadakis2020traffic}). In \cite{delimpaltadakis_tac}, inner approximations $\underline{M}_{\tau_i}$ of IMs $M_{\tau_i}$ were derived as follows: 	
	\begin{equation}\label{inner_approx_eq}
	\underline{M}_{\tau_i} := \{x\in\real^{n}:\mu(x,\tau_i)=0\},
	\end{equation}
	where $\mu(x,\tau_i)$ is a function derived in \cite[Theorem V.3]{delimpaltadakis_tac}.
	Moreover, the sets $\reg_i$ between two approximations $\underline{M}_{\tau_i}$ and $\underline{M}_{\tau_{i+1}}$ (with $\tau_i\leq\tau_{i+1}$) are defined as follows:
	\begin{equation}\label{regs_between_ia_of_manifolds}
	\reg_i = \{x\in\real^n:\mu(x,\tau_i)\leq 0,\mu(x,\tau_{i+1})\geq 0\}
	\end{equation}
	and satisfy:
	\begin{equation}\label{regs_between_ia_of_manifolds_lower_bound}
	\forall x\in \reg_i:\quad \tau(x)\geq\tau_i
	\end{equation}
	
\begin{figure}[!h]
		\centering
		\includegraphics[width=2.4in]{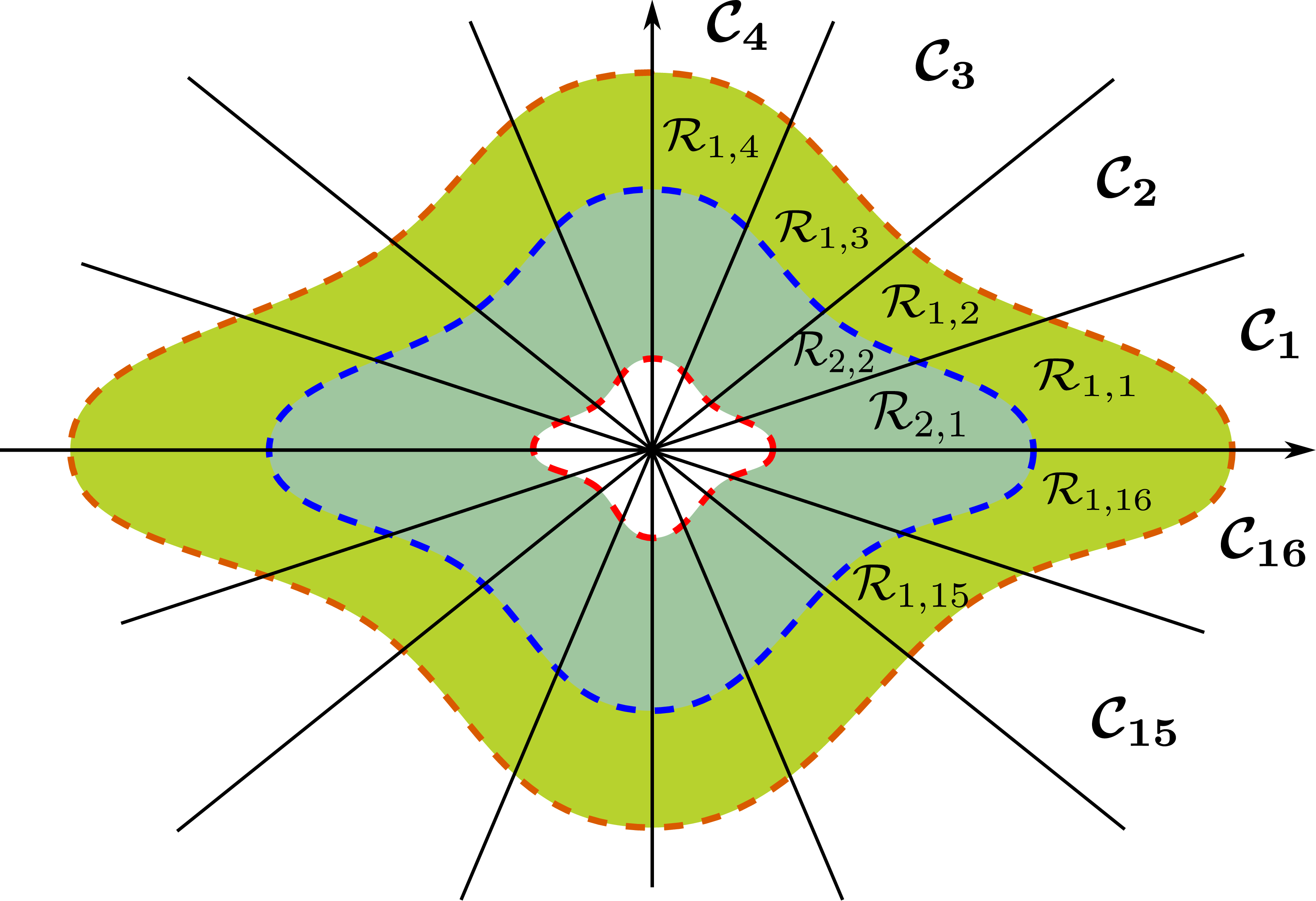}
		\caption{Regions $\reg_{i,j}$ obtained as intersections of inner-approximations of IMs $\underline{M}_{\tau_i}$ (dashed lines) and cones $\cone_j$.}
		\label{fig_partition_cones}
	\end{figure}

	To approximate IM-partitioning, one could divide the set $X$ into such regions $\reg_i$ \eqref{regs_between_ia_of_manifolds}. However, the sets \eqref{regs_between_ia_of_manifolds} are large for the reachability-analysis algorithms of Section \ref{section_reach} to be applied (e.g. see Fig. \ref{fig_two_manifolds}). Thus, we further partition them via cones $\cone_j$ pointed at the origin, which span the whole $\real^n$. Hence, we obtain new sets $\reg_{i,j}$ as intersections of approximations $\underline{M}_{\tau_i}$ and cones $\cone_j$ as follows (see Fig. \ref{fig_partition_cones}):
	\begin{equation}\label{reg_ij}
	\reg_{i,j}=\reg_i\cap\cone_j
	\end{equation}
	Finally, the regions $\tilde{\reg}_{i,j}$ representing the states of the abstraction are obtained as intersections of sets $\reg_{i,j}$ and the set of interest $X$ (the compact state space):
	\begin{equation}\label{tilde_reg_ij}
	\tilde{\reg}_{i,j} = \reg_{i,j}\cap X
	\end{equation} 
	
	To summarize the partitioning method:
	\begin{enumerate}
		\item Define a finite set of times $\{\tau_1,\dots,\tau_q\}$ with $\tau_i<\tau_{i+1}$ and obtain the sets $\reg_i$ according to \eqref{regs_between_ia_of_manifolds}.
		\item Define a conic covering into cones $\cone_j$ and obtain the sets $\reg_{i,j}$ by \eqref{reg_ij}.
		\item Obtain the regions $\tilde{\reg}_{i,j}$ by \eqref{tilde_reg_ij}, which constitute the partition.
	\end{enumerate}
	Note that some regions $\treg_{i,j}$ might be empty sets; such regions are discarded from the abstraction.
	\begin{remark}
		The innermost set $\reg_{q}$ (e.g. the inner white set in Fig. \ref{fig_partition_cones}) is defined: $\reg_q = \{x\in\real^n:\mu(x,\tau_q)\leq 0\}$; $\mu(x,\tau_{q+1})$ is missing, compared to \eqref{regs_between_ia_of_manifolds}, since there is no $\tau_{q+1}$.
	\end{remark}
	\begin{remark}
		As analyzed in \cite{delimpaltadakis_tac} and \cite{delimpaltadakis2020region}, the smaller $\tau_1$ is picked the further away from the origin $\underline{M}_{\tau_1}$ lies. Thus, $\tau_1$ can always be picked such that $X$ is totally covered by the partition. To check if $\underline{M}_{\tau_1}$ is far enough such that $X$ is totally covered, one can employ SMT-solvers (e.g. \cite{dreal}).
	\end{remark}
	\begin{remark}\label{remark_nontight_lower_bound}
		Although \eqref{regs_between_ia_of_manifolds_lower_bound} a-priori provides a lower bound on intersampling times for each region, simulations indicate that the algorithm of Section \ref{section_reach_intervals} often provides less conservative bounds. Nonetheless, with the proposed partitioning method, reachability analysis for timing lower-bounds could be skipped.
	\end{remark}
	\subsection{State Space Partitioning for General Nonlinear Systems}
	Here, we extend the above partitioning method to general nonlinear systems and triggering functions, by employing a homogenization procedure, proposed by \cite{anta2012isochrony}. The homogenization procedure renders an ETC system \eqref{etc_system} and a triggering function homogeneous of degrees $\alpha>0$ and $\theta>0$, respectively, by adding a dummy variable $w$:
	\begin{equation}\label{homogenized_etc_system}
	\begin{aligned}
	\begin{bmatrix}
	\dot{\xi}\\\dot{w}
	\end{bmatrix} &= \begin{bmatrix}
	w^{\alpha+1}f_e(w^{-1}\xi)\\0
	\end{bmatrix} = \tilde{f}_e(\xi,w)\\
	\tilde{\phi}(\xi,w) &= w^{\theta+1}\phi(w^{-1}\xi)
	\end{aligned}
	\end{equation}
	Note that the $\xi$-trajectories of \eqref{homogenized_etc_system} with initial condition $(x_0,e_0,1)\in\real^{2n+1}$ coincide with the trajectories of the ETC system \eqref{etc_system} with initial condition $(x_0,e_0)\in\real^{2n}$; i.e. \eqref{homogenized_etc_system} behaves identically to \eqref{etc_system}, on the $(w=1)$-hyperplane of $\real^{n+1}$. Thus, the state space of the original ETC system \eqref{snh2}-\eqref{trig_cond} (a subset of $\real^n$) is mapped to the $(w=1)$-hyperplane of $\real^{n+1}$.
	
	Employing this procedure, in \cite{delimpaltadakis_tac}, nonlinear ETC systems \eqref{snh2}-\eqref{trig_cond} are homogenized, and then inner-approximations $\underline{M}_{\tau_i}$ of IMs of the homogenized systems \eqref{homogenized_etc_system} are derived in $\real^{n+1}$. These approximations can be used in the same way as in the previous section, to partition the state space. Note that the sets $\reg_{i,j}$ \eqref{reg_ij} are now subsets of $\real^{n+1}$. Since $X$ is now mapped to the set $\{(x,1)\in\real^{n+1}:x\in X\}$, which becomes our set of interest, the regions $\tilde{\reg}_{i,j}$ are now obtained as follows:
	\begin{equation*}
	\tilde{\reg}_{i,j} = \reg_{i,j}\cap\{(x,w)\in\real^{n+1}:x\in X,w=1\},
	\end{equation*}
	\begin{remark}
		As discussed in \cite{delimpaltadakis_tac}, in cases where the origin is the equilibrium of the system and $\phi(0,0)=0$ (e.g. the $\phi$ from \cite{tabuada2007etc}), inner-approximations of IMs exhibit a singularity along the $w$-axis. There is always a small region $\tilde{\reg}_\star$ on the $(w=1)$-hyperplane containing $(0,0,\dots,0,1)$ which is not covered by partitioning with approximations $\underline{M}_{\tau_i}$. $\tilde{\reg}_\star$ can be made arbitrarily small, by choosing $\tau_q$ sufficiently large. Moreover, it can be defined as
		$
		\tilde{\reg}_\star = \{(x,w)\in\real^{n+1}:x\in X,w=1\}\setminus\bigcup\limits_{i,j}\tilde{\reg}_{i,j}
		$
		and treated as an extra state of the abstraction.
	\end{remark}
	\begin{remark}\label{remark_perturbed_partition}
		The proposed partitioning method extends to systems with bounded disturbances/uncertainties, as approximations of IMs of such systems have been derived in \cite{delimpaltadakis2020region}.
	\end{remark}
	\subsection{Discussion}\label{section_partition_discussion}
	Let us discuss the advantages of the proposed partition, compared to naively partitioning $X$ into polytopes. The proposed method is certainly not ideal, as we only have inner approximations of IMs to work with. Nonetheless, our aim was to approximate the ideal IM-partition that was presented in Section \ref{section_IM_partition}, in order to partially gain some of the IM-partition's advantages.
	
	First, the regions $\treg_{i,j}$ generated by the proposed partition are expected to result into tighter intervals, compared to random polytopes of approximately the same volume. That is because they approximate the ideal shape of the regions $R_i$ of Section \ref{section_IM_partition}, which are optimal in terms of correspondence between intersampling interval and volume. This claim is supported by simulation results in Section \ref{section_numerical}, which show that we can partition $X$ with fewer regions \eqref{tilde_reg_ij} than polytopes and still obtain tighter intervals. Hence, with the proposed partition we contain one source of the abstraction's non-determinism.
	
	In addition, due to \eqref{regs_between_ia_of_manifolds_lower_bound}, a region $\treg_{i,j}$ is generated such that $\tau_i$, which is chosen freely, is a lower bound on intersampling times (albeit not the tightest one; see Remark \ref{remark_nontight_lower_bound}). This provides some partial control over the intervals, in contrast to partitioning into random polytopes, where there is no obvious way of relating regions and timing bounds beforehand. Moreover, as a future direction, if outer approximations of IMs were obtained\footnote{Deriving outer approximations of IMs is a difficult problem; e.g. there is no guarantee that a lower bound of the triggering function, derived as in \cite[Lemma V.2]{delimpaltadakis_tac}, exhibits a zero-crossing w.r.t. time for any initial condition.}, they could be used to partition and gain control over the intervals' upper bounds as well (due to the scaling law \cite[Theorem IV.3]{anta2012isochrony}). Finally, the proposed partitioning approach has the potential of approximating arbitrarily well IM-partitioning, by improving the method of approximating IMs.
	
	\section{Numerical Examples}\label{section_numerical}
	Here, we present simulation results supporting our theoretical developments. First, we apply the techniques of Section \ref{section_reach} combined with a naive partition, to abstract a perturbed nonlinear ETC system. Afterwards, we compare the partition proposed in Section \ref{section_partition} with naive partitioning, on an unperturbed system.
	
	In the first example we use Flow*, whereas in the second we use dReach. Moreover, the sets $\reg_{i,j}$ \eqref{reg_ij} are overapproximated by ball segments as described in \cite{delimpaltadakis2020traffic}, as they originally admit a transcendental representation which is currently not supported by either Flow* or dReach. Ball segments can indeed be handled by dReach, but not by Flow*. On the other hand, dReach cannot handle disturbances, but Flow* can. That is why we employ naive partitioning in the perturbed system case. To abstract a perturbed system using the partition of Section \ref{section_partition}, other options have to be explored, such as approximating the sets \eqref{reg_ij} by taylor models, which are handled by Flow*.
	
	To measure the tightness of intervals of a given abstraction, we devise the two following metrics:
		\begin{equation}
		\avgratio = \frac{\sum_{i,j}\frac{\overline{\tau}_{\treg_{i,j}}}{\underline{\tau}_{\treg_{i,j}}}}{\#\mathrm{Regions}}, \quad \avgdiff = \frac{\sum_{i,j}\overline{\tau}_{\treg_{i,j}}-\underline{\tau}_{\treg_{i,j}}}{\#\mathrm{Regions}}
		\end{equation}
		The smaller these metrics are, the tighter the intervals. The difference between them is: in $\avgdiff$ regions with larger intersampling times contribute more to the metric's value, while in $\avgratio$ all regions contribute the same, regardless of the time scales in which they operate. For our purposes, $\avgratio$ is a more representative metric; we have also included $\avgdiff$, since it is closely connected to the formal definition of an abstraction's \textit{precision} (the $\epsilon$ constant from Definition \ref{def_epsilon_approximate}).

	\subsection{Abstracting a Perturbed Nonlinear ETC System}
	Consider the following nonlinear ETC system:
	\begin{equation*}
	\dot{\zeta}_1 = -\zeta_1, \quad
	\dot{\zeta}_2 = \zeta_1^2\zeta_2 + \zeta_2^3 + u + d, \quad \dot{\varepsilon}_{\zeta_1} = -\dot{\zeta}_1, \quad \dot{\varepsilon}_{\zeta_2} = -\dot{\zeta}_2
	\end{equation*}
	with a Lebesgue-sampling triggering function $\phi(\zeta(t),\varepsilon_{\zeta}(t))=\varepsilon_{\zeta}^2-0.01^2$, 
	where $u = -(\zeta_2 + \varepsilon_{\zeta_2}) - (\zeta_1+\varepsilon_{\zeta_1})^2(\zeta_2 + \varepsilon_{\zeta_2}) - (\zeta_2 + \varepsilon_{\zeta_2})^3$ is the control input, and $d\in[-0.1,0.1]$ is a bounded unknown parameter (e.g. a disturbance or a model uncertainty).
	
\begin{figure}[!h]
		\centering
		\includegraphics[width=2.9in]{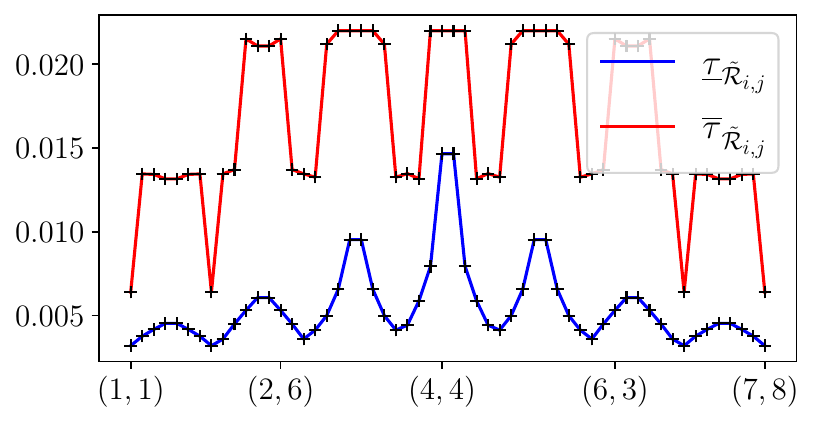}
		\caption{Perturbed ETC System: Timing lower and upper bounds for each region. The horizontal axis shows the regions' indices.}
		\label{perturbed_intervals}
	\end{figure}

\begin{figure}[!h]
		\centering
		\includegraphics[width=2.7in]{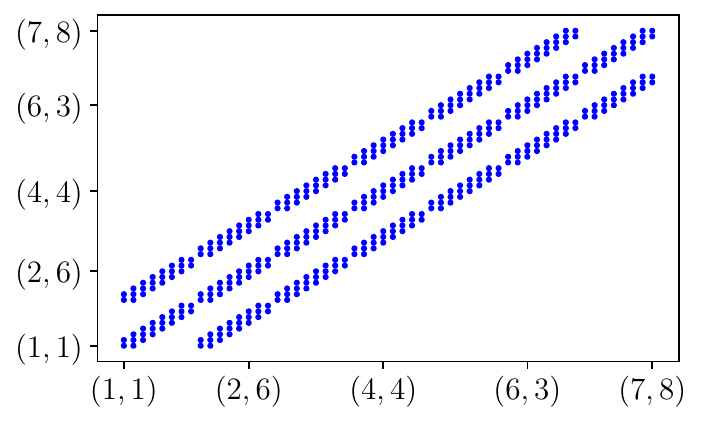}
		\caption{Perturbed ETC System: Transitions of the abstraction. Each dot $[(i,j),(k,l)]$ represents a transition $\treg_{i,j}\to\treg_{k,l}$.}
		\label{example1_transitions}
	\end{figure}

	Let $X=[-2,2]^2$, and we partition it via 56 equal rectangles. We choose a heartbeat $\tau_h = 0.022$. To compute the intervals $[\underline{\tau}_{\treg_{i,j}}, \overline{\tau}_{\treg_{i,j}}]$ and the transitions, we employ the algorithms of Section \ref{section_reach} and Flow*. Figure \ref{perturbed_intervals} depicts the computed timing lower and upper bounds for each region. The tightness metrics are $\avgratio \approx 3.14 $ and $\avgdiff \approx 0.011$. Figure \ref{example1_transitions} depicts the abstraction's transitions (418 in total).
	
\begin{figure}[!h]
		\centering
		\includegraphics[width=3in]{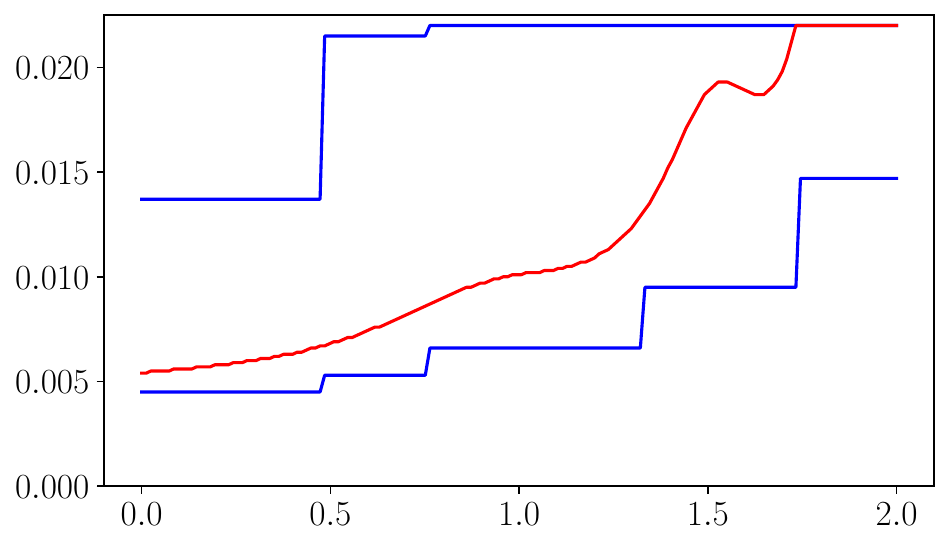}
		\caption{Time evolution of the ETC system's intersampling times (red line) and the intervals $[\underline{\tau}_{\treg_{i,j}},\overline{\tau}_{\treg_{i,j}}]$ (blue lines) generated by the abstraction, during a run.}
		\label{example1_sim}
	\end{figure}

	Finally, we simulate a run of the ETC system to showcase our results' validity. Specifically, the system is initialized at $(1.3,1.3)$, and the disturbance is $d(t)=0.1\sin(10t)$. The duration is $2$s. Figure \ref{example1_sim} depicts the results. The red line is the evolution of the actual ETC intersampling times during the run, while the blue lines represent the intervals $[\underline{\tau}_{\treg_{i,j}},\overline{\tau}_{\treg_{i,j}}]$ generated by the abstraction (by checking at which region $\treg_{i,j}$ the state belonged at each time, and plotting its associated interval). As expected, the intersampling time is always confined in $[\underline{\tau}_{\treg_{i,j}},\overline{\tau}_{\treg_{i,j}}]$. Moreover, it caps at $\tau_h=0.022$. The system's trajectory followed the spatial path: $\treg_{6,7}\to\dots\to\treg_{6,6}\to\dots\to \treg_{5,6}\to\dots\to\treg_{5,5}\to\dots\to \treg_{4,5}\to\dots$, where the dots indicate that the trajectory stayed in the previous region for multiple intersampling intervals. Note that all transitions taken during the run are contained in the transition set of the abstraction (Fig. \ref{example1_transitions}).
	\subsection{Performance of the Partitioning Approach of Section \ref{section_partition}}
\begin{figure}[!h]
		\centering
		\includegraphics[width=2.9in]{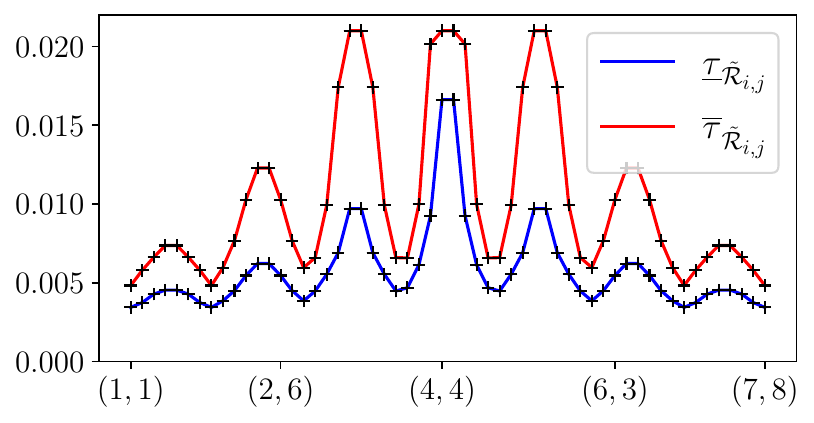}
		\caption{Naive Partition: Lower and upper bounds of intersampling times for each region.}
		\label{naive_regions_times}
	\end{figure}

	To compare our proposed partition with naive partitioning, consider the unperturbed version of the ETC system presented in the previous numerical example, let $X=[-2,2]^2$ and $\tau_h = 0.021$. For naive partitioning, we divide again $X$ into 56 equal rectangles and calculate the intervals $[\underline{\tau}_{\treg_{i,j}}, \overline{\tau}_{\treg_{i,j}}]$. The results appear in Fig. \ref{naive_regions_times}. The tightness metrics are $\avgratio \approx 1.74$ and $\avgdiff \approx 0.0045$. Additionally, the total transitions of the abstraction are 367. We observe that the timing intervals are considerably tighter and the number of transitions is smaller, when the disturbance is absent. That is because unknown parameters in the dynamics give rise to infinite possible behaviours, implying larger non-determinism. Moreover, reachability-analysis tools overapproximate more conservatively, when unknown parameters are present.
	
\begin{figure}[!h]
		\centering
		\includegraphics[width=2.9in]{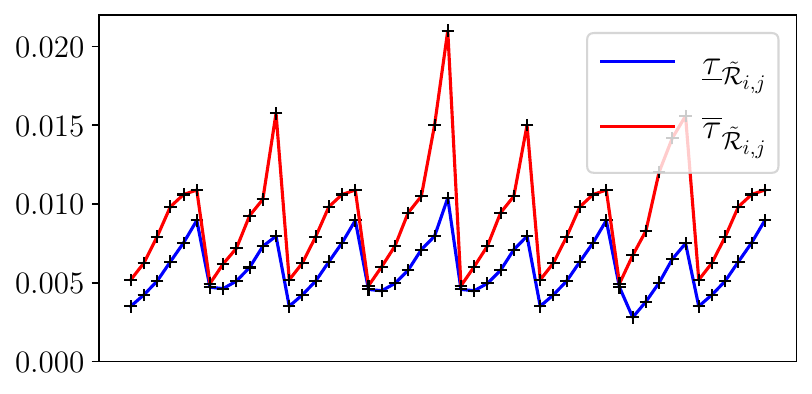}
		\caption{Proposed Partition: Lower and upper bounds of intersampling times for each region.}
		\label{impartition_regions_times}
	\end{figure}

	For the partitioning approach of Section \ref{section_partition}, after homogenizing the system and the triggering function as in \eqref{homogenized_etc_system} with $\alpha = 2$ and $\theta = 1$, we define the set of times $\{.002, .0028, .0038, .005, .0065, .0075\}$ and derive inner-approximations of the corresponding IMs and the sets $\reg_i$, as per \cite{delimpaltadakis_tac} and \eqref{regs_between_ia_of_manifolds}. To further divide $\reg_i$ into $\reg_{i,j}$, we use 9 polyhedral cones $\cone_j$ pointed at the origin of $\real^{n+1}$, that cover the set of interest $\{(x,w)\in\real^{n+1}:x\in X,w=1\}$; i.e. $\bigcup_j(\cone_j\cap\{(x,w)\in\real^{n+1}:w=1\}) = \{(x,w)\in\real^{n+1}:x\in X,w=1\}$\footnote{A way to create this conic covering is to divide $\{(x,w):x\in X,w=1\}$ into 9 squares, and obtain $\cone_j$ as the conic hull of the $j$-th square's vertices.}. Finally, after obtaining the regions $\treg_{i,j}$ \eqref{tilde_reg_ij}, the total number of abstraction states is 49 (recall that the number of regions $\treg_{i,j}$ can be smaller than $|\{\reg_i\}|\cdot|\{\cone_j\}|$, where $|\cdot|$ denotes set cardinality, since empty intersections \eqref{tilde_reg_ij} are discarded). The computed intervals $[\underline{\tau}_{\treg_{i,j}}, \overline{\tau}_{\treg_{i,j}}]$ are depicted in Fig. \ref{impartition_regions_times}. The tightness metrics are $\avgratio \approx 1.54$ and $\avgdiff \approx 0.0032$. The total number of transitions is 471.
	
	The partition of Section \ref{section_partition} achieves considerably tighter intervals even with a smaller amount of regions, compared to the naive one. This supports the claims of Section \ref{section_partition_discussion}: it leads to tighter intervals, thus containing one of the sources of non-determinism. On the other hand, we observe that it has led to an abstraction with larger transition set. That may be because the sets \eqref{reg_ij} have been overapproximated by ball segments, which in some cases might be a crude approximation (see \cite{delimpaltadakis2020traffic}), while the naive partition's rectangles are fed directly to the reachability-analysis algorithm. In other words, while tighter intervals are an inherent characteristic of the partition of Section \ref{section_partition}, the large number of transitions is probably due to coarse overapproximations.
	
	\section{Conclusion and Future Work}\label{section_conclusion}
	We  constructed traffic abstractions of perturbed uncertain nonlinear ETC systems with general triggering functions. Thus, we have significantly extended the applicability of abstraction-based scheduling of traffic in networks of ETC loops, which was only applicable to LTI systems with quadratic triggering functions so far. To capture the sets of intersampling times that the given ETC system may generate, we formulated and solved reachability-analysis problems. In addition, we proposed a state-space partitioning based on IMs, which provides partial control over the abstraction's accuracy and leads to tighter timing intervals, compared to naive partitioning. However, in the performed simulations it has lead to larger transition sets, probably because of the crude overapproximations used to facilitate reachability analysis. In future work, we plan to: a) perform experiments showcasing abstraction-based scheduling on networks of nonlinear ETC systems, b) develop more accurate approximations of the sets \eqref{reg_ij} (e.g. polynomial zonotopes or taylor models), to reduce the size of the transition set, while keeping the timing intervals tight, thus overall containing the abstraction's non-determinism, and c) employ the derived abstractions to characterize the sampling performance of ETC (e.g. compute performance metrics of ETC systems, as done in \cite{gleizer2021computing} for the \textit{minimum average intersampling time}).
	
	\bibliography{nonlinear_abs_bib.bib}

\begin{thebibliography}{10}
\providecommand{\url}[1]{#1}
\csname url@samestyle\endcsname
\providecommand{\newblock}{\relax}
\providecommand{\bibinfo}[2]{#2}
\providecommand{\BIBentrySTDinterwordspacing}{\spaceskip=0pt\relax}
\providecommand{\BIBentryALTinterwordstretchfactor}{4}
\providecommand{\BIBentryALTinterwordspacing}{\spaceskip=\fontdimen2\font plus
\BIBentryALTinterwordstretchfactor\fontdimen3\font minus
  \fontdimen4\font\relax}
\providecommand{\BIBforeignlanguage}[2]{{%
\expandafter\ifx\csname l@#1\endcsname\relax
\typeout{** WARNING: IEEEtran.bst: No hyphenation pattern has been}%
\typeout{** loaded for the language `#1'. Using the pattern for}%
\typeout{** the default language instead.}%
\else
\language=\csname l@#1\endcsname
\fi
#2}}
\providecommand{\BIBdecl}{\relax}
\BIBdecl

\bibitem{astrom2002comparison}
K.~J. Astrom and B.~M. Bernhardsson, ``Comparison of riemann and lebesgue
  sampling for first order stochastic systems,'' in \emph{Proceedings of the
  41st IEEE Conference on Decision and Control, 2002.}, vol.~2.\hskip 1em plus
  0.5em minus 0.4em\relax IEEE, 2002, pp. 2011--2016.

\bibitem{tabuada2007etc}
P.~Tabuada, ``Event-triggered real-time scheduling of stabilizing control
  tasks,'' \emph{IEEE Transactions on Automatic Control}, vol.~52, no.~9, pp.
  1680--1685, 2007.

\bibitem{small_gain_robust_etc}
T.~Liu and Z.-P. Jiang, ``A small-gain approach to robust event-triggered
  control of nonlinear systems,'' \emph{IEEE Transactions on Automatic
  Control}, vol.~60, no.~8, pp. 2072--2085, 2015.

\bibitem{girard2015dynamicetc}
A.~Girard, ``Dynamic triggering mechanisms for event-triggered control,''
  \emph{IEEE Transactions on Automatic Control}, vol.~60, no.~7, pp.
  1992--1997, 2015.

\bibitem{2012introtoetc_stc}
W.~P. M.~H. Heemels, K.~H. Johansson, and P.~Tabuada, ``An introduction to
  event-triggered and self-triggered control,'' in \emph{Proceedings of the
  IEEE Conference on Decision and Control}, 2012, pp. 3270--3285.

\bibitem{buttazzo1998elastic}
G.~C. Buttazzo, G.~Lipari, and L.~Abeni, ``Elastic task model for adaptive rate
  control,'' in \emph{Proceedings 19th IEEE Real-Time Systems Symposium (Cat.
  No. 98CB36279)}.\hskip 1em plus 0.5em minus 0.4em\relax IEEE, 1998, pp.
  286--295.

\bibitem{caccamo2000elastic}
M.~Caccamo, G.~Buttazzo, and L.~Sha, ``Elastic feedback control,'' in
  \emph{Proceedings 12th Euromicro Conference on Real-Time Systems. Euromicro
  RTS 2000}.\hskip 1em plus 0.5em minus 0.4em\relax IEEE, 2000, pp. 121--128.

\bibitem{bhattacharya2004anytime}
R.~Bhattacharya and G.~J. Balas, ``Anytime control algorithm: Model reduction
  approach,'' \emph{Journal of Guidance, Control, and Dynamics}, vol.~27,
  no.~5, pp. 767--776, 2004.

\bibitem{fontanelli2008scheduling}
D.~Fontanelli, L.~Greco, and A.~Bicchi, ``Anytime control algorithms for
  embedded real-time systems,'' \emph{Lecture Notes in Computer Science
  (including subseries Lecture Notes in Artificial Intelligence and Lecture
  Notes in Bioinformatics)}, vol. 4981 LNCS, pp. 158--171, 2008.

\bibitem{areqi2015scheduling}
S.~Al-Areqi, D.~Görges, and S.~Liu, ``Event-based networked control and
  scheduling codesign with guaranteed performance,'' \emph{Automatica},
  vol.~57, pp. 128--134, 2015.

\bibitem{lu2002feedback}
C.~Lu, J.~A. Stankovic, S.~H. Son, and G.~Tao, ``Feedback control real-time
  scheduling: Framework, modeling, and algorithms,'' \emph{Real-Time Systems},
  vol.~23, no. 1-2, pp. 85--126, 2002.

\bibitem{cervin2005control}
A.~Cervin and J.~Eker, ``Control-scheduling codesign of real-time systems: The
  control server approach,'' \emph{Journal of Embedded Computing}, vol.~1,
  no.~2, pp. 209--224, 2005.

\bibitem{arman_formal_etc}
A.~S. Kolarijani and M.~Mazo~Jr., ``Formal traffic characterization of lti
  event-triggered control systems,'' \emph{IEEE Transactions on Control of
  Network Systems}, vol.~5, no.~1, pp. 274--283, 2016.

\bibitem{gleizer2020scalable}
G.~de~A.~Gleizer and M.~Mazo~Jr., ``Scalable traffic models for scheduling of
  linear periodic event-triggered controllers,'' \emph{IFAC-PapersOnLine},
  vol.~53, no.~2, pp. 2726--2732, 2020, 21st IFAC World Congress.

\bibitem{dreach}
S.~Kong, S.~Gao, W.~Chen, and E.~Clarke, ``dreach: $\delta$-reachability
  analysis for hybrid systems,'' in \emph{International Conference on TOOLS and
  Algorithms for the Construction and Analysis of Systems}.\hskip 1em plus
  0.5em minus 0.4em\relax Springer, 2015, pp. 200--205.

\bibitem{flowstar}
X.~Chen, E.~{\'A}brah{\'a}m, and S.~Sankaranarayanan, ``Flow*: An analyzer for
  non-linear hybrid systems,'' in \emph{International Conference on Computer
  Aided Verification}.\hskip 1em plus 0.5em minus 0.4em\relax Springer, 2013,
  pp. 258--263.

\bibitem{delimpaltadakis_tac}
G.~Delimpaltadakis and M.~Mazo~Jr., ``Isochronous partitions for region-based
  self-triggered control,'' \emph{IEEE Transactions on Automatic Control},
  vol.~66, no.~3, pp. 1160--1173, 2021. doi: 10.1109/TAC.2020.2994020

\bibitem{delimpaltadakis2020region}
------, ``Region-based self-triggered control for perturbed and uncertain
  nonlinear systems,'' \emph{IEEE Transactions on Control of Network Systems},
  2021. doi: 10.1109/TCNS.2021.3050121

\bibitem{delimpaltadakis2020traffic}
------, ``Traffic abstractions of nonlinear homogeneous event-triggered control
  systems,'' in \emph{2020 59th IEEE Conference on Decision and Control
  (CDC)}.\hskip 1em plus 0.5em minus 0.4em\relax IEEE, 2020, pp. 4991--4998.

\bibitem{tabuada_book}
P.~Tabuada, \emph{Verification and control of hybrid systems: a symbolic
  approach}.\hskip 1em plus 0.5em minus 0.4em\relax Springer Science \&
  Business Media, 2009.

\bibitem{dreal}
S.~Gao, S.~Kong, and E.~M. Clarke, ``dreal: An smt solver for nonlinear
  theories over the reals,'' in \emph{International Conference on Automated
  Deduction}.\hskip 1em plus 0.5em minus 0.4em\relax Springer, 2013, pp.
  208--214.

\bibitem{anta2012isochrony}
{A. Anta and P. Tabuada}, ``Exploiting isochrony in self-triggered control,''
  \emph{IEEE Transactions on Automatic Control}, vol.~57, no.~4, pp. 950--962,
  2012.

\bibitem{gleizer2021computing}
G.~de~A.~Gleizer and M.~Mazo~Jr., ``Computing the sampling performance of
  event-triggered control,'' in \emph{Proceedings of the 24th International
  Conference on Hybrid Systems: Computation and Control}, 2021, pp. 1--7.

\end{thebibliography}
	\bibliographystyle{myIEEEtran}
\end{document}